# Supervised Classifiers for Audio Impairments with Noisy Labels

*Chandan K A Reddy, Ross Cutler, Johannes Gehrke*

Microsoft Corporation, Redmond WA

{chandan.karadagur, rcutler, johannes}@microsoft.com

## Abstract

Voice-over-Internet-Protocol (VoIP) calls are prone to various speech impairments due to environmental and network conditions resulting in bad user experience. A reliable audio impairment classifier helps to identify the cause for bad audio quality. The user feedback after the call can act as the ground truth labels for training a supervised classifier on a large audio dataset. However, the labels are noisy as most of the users lack the expertise to precisely articulate the impairment in the perceived speech. In this paper, we analyze the effects of massive noise in labels in training dense networks and Convolutional Neural Networks (CNN) using engineered features, spectrograms and raw audio samples as inputs. We demonstrate that CNN can generalize better on the training data with a large number of noisy labels and gives remarkably higher test performance. The classifiers were trained both on randomly generated label noise and the label noise introduced by human errors. We also show that training with noisy labels requires a significant increase in the training dataset size, which is in proportion to the amount of noise in the labels.

## 1. Introduction

In recent times, the quality of speech transmitted over the Internet is being monitored closely by most of the voice service providers as it correlates highly with the user experience. The speech signal perceived by the human is degraded due to various environmental noises, bad room acoustics and distortions introduced in the communication systems. The environmental noises include a variety of background noises and reverberation due to bad acoustics of the room. The communication system introduces speech distortion due to bad microphones, poor network conditions and the audio processing components such as noise suppression and a gain controller. The poor network conditions will result in the packet loss concealment and the distortion of the speech perceived. Over the years, the International Telecommunication Union-Telephony (ITU-T) proposed various standards to quantify the quality of the audio impaired due to various reasons. ITU-T Recommendation P.800 [1], a subjective evaluation procedure using Mean Opinion Score (MOS) is the gold standard for quantifying the speech quality. Objective speech quality metrics such as Perceptual Evaluation of Speech Quality (PESQ) [2], Perceptual objective listening quality prediction (POLQA) [3] and [4] are designed to predict MOS. The quality score is useful in knowing the overall performance of the communication system, but it does not give us the reason for bad speech quality when the quality score is lower. Having an audio impairment classifier can precisely point us to the problem areas, thereby aiding the engineers to take actions and fix the problems. For example, if the audio impairment classifier detects speech distortion due to a bad network, the network parameters can be tuned and optimized to work in low bandwidth conditions.

There is an increasing demand for large datasets to train deep architectures to achieve high classification accuracy. The data used for training a supervised classifier should cover enough variety to capture most of the realistic scenarios, and at the same time should have higher quality labels to train the classifier. For training an audio impairment classifier, the audio data can be synthesized to emulate various impairment types and hence there is no requirement for annotating the clips as we know the ground truth labels. However, synthetic data restricts the space of impairments that can be addressed. An alternative solution is to capture the data from realistic scenarios using the actual microphones and variety of acoustic and network conditions. However, the data captured in the wild need's human labels on the type of impairments perceived. The reliability of these labels depends on the expertise of the human labelers, which may be unattainable at scale. The labels obtained from non-expert listeners using online platforms such as Microsoft's Universal Human Relevance System (UHRS) or Amazon's Mechanical Turk [5] are noisy and less reliable. Unsupervised and self-supervised learning would be reasonable solutions when there are no labels available. Even though we have labels with < 100% accuracy they still overall have useful information and it makes sense to use them during training. Recently, research has shown that Deep Neural Networks (DNN) generalize well even after training with large noisy labeled data [6], [7]. Specifically, Convolutional Neural Networks (CNNs) are shown to generalize well for image classification problems [8]. While the research is extensive on noisy labels in the field of computer vision, there is little work published in the field of audio and speech processing, which is mainly in audio event detection [9], [10], [11].

In this work, we study three supervised classifiers for audio impairments under low label reliability. We explore three approaches to train a neural network with audio data with noisy labels. In the first approach, hand-engineered signal processing-based features are used as input to train a dense network. In the second approach, we use Log Mel Spectrogram as input to train a 2D-CNN. Finally, we use raw audio samples to train a 1D-CNN for classifying impairments. The insights from our study show that CNN architectures can generalize well even in the presence of arbitrarily large label noise. The classifier was able to achieve an accuracy of about 90%, despite being trained on labels at 80% error rate. This work, to our knowledge, is the first time an audio impairment classifier for a voice calling application is trained on noisy labels due to human errors.

## 2. Audio dataset

In this project, we considered 4 impairment classes and 1 no-impairment class. The 4 impairment classes are: i) Background noise, ii) Reverberation, iii) Speech distortion and iv) Low volume. These are some of the top audio impairments that users perceive frequently in Skype VoIP calls. The no-impairment class is composed of clean speech dataset. We created a dataset of 5000 audio clips uniformly distributed across impairment types with an average clip length of 10 seconds. The 1000 clean speech clips were equally distributed by gender: 500 male and 500 female speakers. The background noise clips captured different noisy environments such as office, air conditioner, car, restaurant, and cafeteria noises. The noise was added to clean speech at different Speech to Noise Ratio (SNR) levels. The SNR levels of 0 to 30 dB with 5 dB increments were considered. Audio clips with reverberation were synthesized by convolving the clean speech with room impulse responses with $RT_{60}$ ranging from 300 to 1200 ms with 300 ms increments. The speech distortion due to network conditions was simulated using the Test Authoring and Execution Framework (TAEF) [12] using the network traces captured from actual Skype voice calls. The trace contains information about the packet loss concealment across time. The TAEF tool is used to apply the trace on any audio clip to synthesize its distorted version. The traces that give a Mean Opinion Score (MOS) degradation between 1 and 2 was applied on clean speech dataset to generate distorted speech clips. The low volume clips were synthesized by scaling the clean speech data to have Root Mean Square (RMS) levels in the range -50 dBFS to -35 dBFS. All the clips except low volume are normalized to -25 dBFS for playback and are sampled at the rate of 16 kHz.

## 3. Online evaluation and noisy labels

### 3.1. Hit application for online subjective evaluation

Once the audio dataset is synthesized for different impairments, the next step is to label the clips. Note that the ground truth labels are known since the data is synthesized. Nevertheless, we collect labels from human judges to capture their noise. The synthesized audio dataset with 5000 clips is about 14 hours long. Annotating these clips offline with 10 expert listeners per clip is not practically feasible as it is expensive and time-consuming. Online tools such as UHRS and Mechanical Turk are used extensively in recent times to get annotations and labels for datasets used in training machine learning models for various applications. These tools are cost-effective as the tasks are crowdsourced online to thousands of click workers across the world. These individuals are compensated based on the number of jobs or hits they complete. Since these tasks can be completed simultaneously by many click workers, the process of getting labels is extremely fast compared to offline tests. The downside of using these tools is that the click workers are non-expert listeners and we have very little control over the quality of their judgments. This is very similar to the case of collecting the user feedback about the quality after every voice call.

We used Microsoft's UHRS tool for our experiments. UHRS is very similar to Mechanical Turk, but internal to Microsoft. Experiments conducted using UHRS can be easily replicated using Mechanical Turk, which is publicly available. The Hit application designed is shown in Figure 1. For quality control, the click workers had to go through a training phase during which, they listened to about 8 clips that covered different types of impairments. This is to get them familiar with the kind of impairments they will be tested upon during the actual experiments. The training is followed by the qualification test in which the raters are given audio clips that are either extremely noisy or clean speech. So, the expected rating is either 1 or 5. This will ensure that the click workers have a playback device and that they are not hearing impaired. If the workers pass the qualification test, they will proceed to the main experiments where they can click on each audio clip and rate the quality and click on the impairments they perceived. Every audio clip was rated by 10 judges and in total, we had 50000 judgments for 5000 clips.

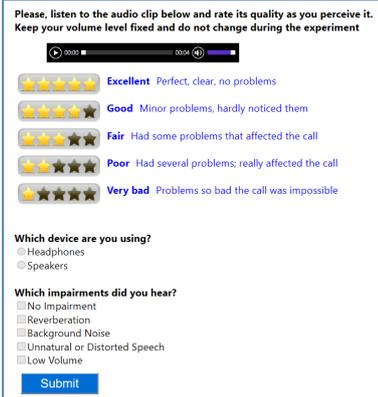

Figure 1: UHRS Hit Application

### 3.2. Characteristics of noisy labels

In this work, we use labels with both synthetic noise and real label noise from UHRS human judgments. The label noise literature typically deals with synthetic noise [13], [14], [15]. The synthetic noise is usually drawn from a random distribution. Synthetic noise gives us the ability to generate massive amounts of noisy labels in the dataset. We consider the cases in which the label error rate is 20%, 50%, and 80% uniformly distributed across labels. We also considered the case in which we had 2, 4 and 8 noisy labels for every clean label. In every case, there were non-zero and enough clean labels. The literature shows that it is more destructive to have a smaller number of clean labels than having a greater number of noisy labels [16]. The labels from UHRS experiments were 66% accurate with the noise having a stochastic distribution. Note that the noise in the labels is despite the training and qualification that the UHRS click workers went through. The actual call feedback from users will be much noisier.

## 4. Supervised classifiers

### 4.1. Engineered audio features with dense network

In the first approach, we extract 18 engineered signal processing features from the audio signal. The features extracted are spectral centroid, spectral flux, spectral flatness, spectral dynamics, spectral roll-off, zero crossing rate, signal energy, energy entropy, Global SNR and clipping probability [17]. We compute mean and variance across the entire clip length for all the features except SNR and clipping probability, which sums up to 18 features. Each of these features captures different characteristics of the audio signal such as stationarity of the noise, clipping, tracking high-frequency noise components, reverb, speech distortion, etc. These engineered features are used as input to train a Fully Connected (FC) network for classification. Figure 2 (a) shows the network architecture consisting of 3 fully connected layers with a dropout rate of 0.5 and Rectified Linear Unit (ReLU) activations. The advantage of using handcrafted features is that the input dimension to a classifier is greatly reduced and hence requires a smaller network for classification. The downside of using engineered features is that we are limited by our signal processing knowledge to capture the properties of all the impairments. This approach is not scalable to accommodate newer and unseen impairment types.

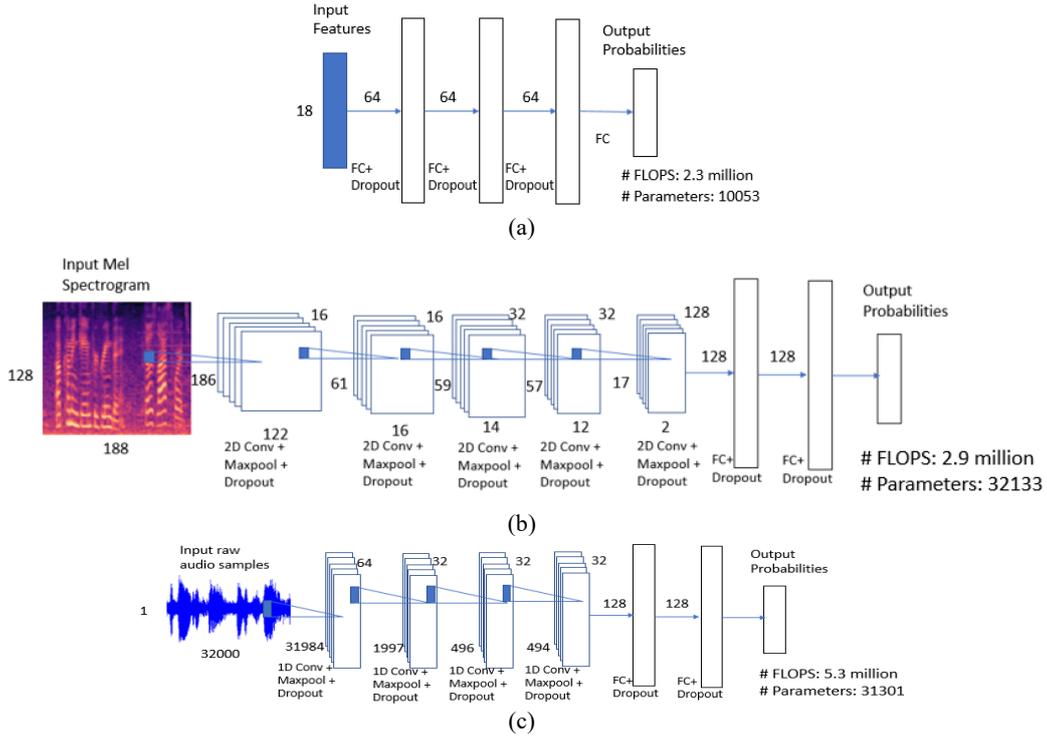

Figure 2: Block diagrams of audio impairment classifiers using (a) Engineered features, (b) Mel Spectrogram and 2D CNN and (c) Raw Audio Samples and 1D CNN

### 4.2. Log mel spectrogram features

In the second approach, the Log Mel Spectrogram is used as input to train the classifier. The audio signal is processed in frames of 320 samples (20 ms) with a hop size of 160 samples and 128 Mel bands were generated per frame. The Mel features were extracted for 3 secs of audio, which constitute 188 frames. Hence, the input spectrogram of size 128 x 188 is used to train a 2D-CNN. The network architecture details are in Figure 2 (b). The CNN layers are followed by max-pooling layer to reduce the dimensionality. The dropout rate of 0.1 was used. Two FC layers with 128 hidden nodes follow the CNN block and ReLU is used as activation throughout. The reason for choosing 3 secs is because anything less is too short to capture the impairments, especially when the speech distortion is due to packet loss and there are cases when the entire 3 secs clip is null. The Mel spectrogram is extracted from the magnitude spectrum of the signal which captures key information about the impairments. However, it ignores the information from the phase. There is a mixed response from the speech community about the importance of phase on perceptual quality. The classifier is computationally fast with a number of Flops of about 2.9 million.

### 4.3. Raw audio samples as input

The features described in the previous subsections are limited by our signal processing knowledge and the transformations applied. The spectrogram eliminates the phase spectral information from the signal, which is shown to have an impact on human perception. In this section, we will use the raw audio samples as input to the 1D-CNN architecture shown in Figure 2 (c). When using raw audio samples, there is no loss of information due to human knowledge. We allow the neural network to extract the relevant information and learn the task. 2 secs of raw audio constitute 32000 samples at 16 kHz sampling rate. Unless, processed frame wise, the biggest drawback of using raw audio samples is its large input dimension, which results in bigger network architecture. Every 1D-CNN layer is followed by a Max Pooling layer dropout rate of 0.1 is used. The last two layers are FC with 128 hidden nodes each. ReLU is used throughout the network.

## 5. Experiments and results

### 5.1. Baseline evaluation

The three supervised classifiers described in Section 4 are trained with clean labels and the evaluation results are used as the baseline to analyze the impact on the accuracy when trained with erroneous labels. The data is divided into 70% for training, 15% for validation and testing each. The engineered features and the Log Mel Spectrogram described in sections 4.1 and 4.2 are computed on 3-sec audio signals that are randomly sampled from the 10-sec audio clips from the training dataset. Similarly, 2-sec segments of raw audio data are sampled randomly from the training dataset for the approach described in 4.3. The training of these models was continued until the models converged. Hence, the amount of data required for training varied for each of these methods. Table 1 shows the classification accuracy for the three Supervised Classifiers described in Section 4. All the validation and testing are done on the clean labels. The training set size column gives the number of training examples required for the loss to converge. The error rate of 0% corresponds to training on the clean labels. The methods in Figures 2 (b) and 2 (c) with Log Mel Spectrogram as input to 2D CNN and raw audio as input to 1D CNN performs equally well when trained on clean labels. The accuracy is acceptable for the case of engineered features and FC layers, given that the setup is computationally efficient.

Table 1: *Classification Accuracy on clean and noisy labels for different approaches*

| Error Rate | Engineered features | | Log Mel Spectrogram+2D CNN | | Raw Audio Input + 1D CNN | |
|---|---|---|---|---|---|---|
| | Accuracy | Training set size | Accuracy | Training set size | Accuracy | Training set size |
| **0 %** | 91% | 10000 | 99% | 100000 | 98% | 100000 |
| **20%** | 72% | 12000 | 96% | 150000 | 91% | 150000 |
| **50%** | 49% | 13000 | 93% | 400000 | 89% | 400000 |
| **80%** | 35% | 14000 | 89% | 500000 | 77% | 500000 |
| **2 noisy per clean** | 56% | 13000 | 92% | 500000 | 87% | 500000 |
| **8 noisy per clean** | 43% | 15000 | 83% | 600000 | 74% | 600000 |
| **UHRS labels (34% erroneous)** | 68% | 12000 | 91% | 100000 | 89% | 100000 |

### 5.2. Classification accuracy when trained on noisy labels

The error rates of 20%, 50%, and 80% correspond to a decrease in the number of clean labels assuming a fixed dataset size. The other two conditions with 2 and 8 noisy labels for every clean label correspond to increasing the number of noisy labels without decreasing the number of clean labels. For example, assuming the original dataset has 10000 clean labels, in the case of 80% error rate, 8000 of those labels are randomized to make them noisy. For the case of 2 noisy labels for every clean label, we keep the 10000 clean labels and synthesize 20000 noisy labels and append it with the training dataset. The final condition is the noise in the labels due to human judgments from UHRS. The results in Table 1 shows that the accuracy decreases when trained on noisy labels. But the drop in the accuracy is least for Log Mel Spectrogram with 2D CNN and highest for the case of using 18 engineered features. Our experimental results are in accord with other computer vision literature work showing that the 2D CNN architectures generalize well even for audio impairment classification when trained on massive synthetic label noise and on the real noise due to human judgments. The confusion matrix for Log Mel Spectrogram and 2D CNN trained on UHRS noisy labels is shown in Table 2. This case is of primary interest for the audio impairment classification when trained on user feedback as labels. The higher misclassification rate for 'No Impairment' case is because there were many examples where human judges selected 'No Impairment' when there was background noise or low volume.

### 5.3. Relationship between training set size and label noise

The amount of training data with noisy labels required to attain a certain accuracy increases with an increase in the number of noisy labels in the dataset. The critical number of clean labels required depends on the amount of noise in the labels. The requisite of clean labels increases non-linearly with the ratio of noisy to clean examples. The accuracy threshold is fixed to 85% and the training data required to achieve the accuracy with an increase in the label noise is monitored. Figure 3 shows that the requirement of training examples increases exponentially with an increase in the label noise.

Table 2: *Confusion matrix for training the Log Mel Spectrogram + 2D CNN with noisy labels*

| | | Ground truth | | | | |
|---|---|---|---|---|---|---|
| | | Background Noise | Speech Distortion | Reverb | Low Volume | No Impairment |
| **Predicted** | Background Noise | **0.997** | 0.089 | 0.000 | 0.000 | 0.091 |
| | Speech Distortion | 0.002 | **0.911** | 0.059 | 0.010 | 0.000 |
| | Reverb | 0.000 | 0.000 | **0.941** | 0.000 | 0.000 |
| | Low Volume | 0.000 | 0.000 | 0.000 | **0.936** | 0.082 |
| | No Impairment | 0.000 | 0.000 | 0.000 | 0.050 | **0.827** |

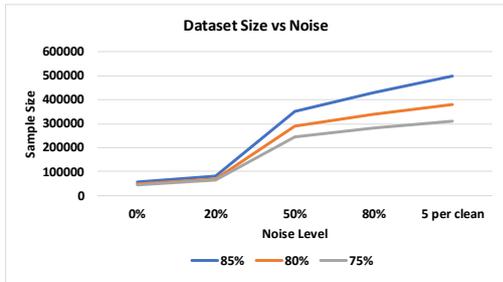

Figure 3: Relationship between training dataset size and noise levels for Accuracy threshold of 75%, 80% and 85%

## 6. Conclusion

In this paper, we investigated the effects of noisy labels on training an audio impairment classifier using three different input and network architectures. Experimental results suggest that a Log Mel Spectrogram with 2D CNN architecture can be a feasible option to train a supervised audio impairment classifier with noisy labels, provided a sufficient number of clean labels are available. Despite human labels are only 66% accurate, the accuracy of the classifier on an independent test set does not degrade. This robustness can dramatically reduce false positives that is invaluable in live systems running at scale. It is worth to note that the performance also depends on the distribution of the noise in the labels.

# 7. References

[1] ITU-T, "Recommendation P.800: Methods for subjective determination of transmission quality," Feb. 1998.

[2] ITU-T, "Recommendation P.862: Perceptual evaluation of speech quality (PESQ), an objective method for end-to-end speech quality assessment of narrowband telephone networks and speech codecs," Feb. 2001.

[3] J. G. Beerends et al., "Perceptual objective listening quality assessment (POLQA), the third generation ITU-T standard for end-to-end speech quality measurement part I—Temporal alignment," J. Audio Eng. Soc., vol. 61, no. 6, pp. 366–384, 2013.

[4] A. R. Avila, H. Gamper, C. Reddy, R. Cutler, I. Tashev and J. Gehrke, "Non-intrusive Speech Quality Assessment Using Neural Networks," *ICASSP 2019 - 2019 IEEE International Conference on Acoustics, Speech and Signal Processing (ICASSP)*, Brighton, United Kingdom, 2019, pp. 631-635. doi: 10.1109/ICASSP.2019.8683175

[5] https://www.mturk.com/

[6] N. Natarajan, I. S. Dhillon, P. K. Ravikumar, and A. Tewari, "Learning with noisy labels," in Advances in neural information processing systems, 2013, pp. 1196–1204.

[7] Jindal, Ishan, Matthew S. Nokleby and Xuewen Chen. "Learning Deep Networks from Noisy Labels with Dropout Regularization." *2016 IEEE 16th International Conference on Data Mining (ICDM)* (2016): 967-972.

[8] Sukhbaatar, Sainbayar and Rob Fergus. "Learning from Noisy Labels with Deep Neural Networks." *CoRR* abs/1406.2080 (2014): n. pag.

[9] Fonseca, Eduardo, Manoj Plakal, Daniel P. W. Ellis, Frederic Font, Xavier Favory and Xavier Serra. "Learning Sound Event Classifiers from Web Audio with Noisy Labels." *CoRR* abs/1901.01189 (2019): n. pag.

[10] Iqbal, Turab, Qiuqiang Kong, Mark. Plumbley and Wenwu Wang. "General-purpose audio tagging from noisy labels using convolutional neural networks." (2018).

[11] Shah, Ankit, Anurag Kumar, Alexander G. Hauptmann and Bhiksha Raj. "A Closer Look at Weak Label Learning for Audio Events." *CoRR* abs/1804.09288 (2018): n. pag.

[12] https://docs.microsoft.com/en-us/windows-hardware/drivers/taef/

[13] Scott E. Reed, Honglak Lee, Dragomir Anguelov, Christian Szegedy, Dumitru Erhan, and Andrew Rabinovich, "Training deep neural networks on noisy labels with bootstrapping," in ICLR 2015.

[14] Daiki Tanaka, Daiki Ikami, Toshihiko Yamasaki, and Kiyoharu Aizawa, "Joint optimization framework for learning with noisy labels," in Proceedings of CVPR, 2018, pp. 5552–5560.

[15] Zhilu Zhang and Mert Sabuncu, "Generalized cross entropy loss for training deep neural networks with noisy labels," in Advances in Neural Information Processing Systems, 2018.

[16] D. Rolnick, A. Veit, S. J. Belongie, and N. Shavit. Deep learning is robust to massive label noise. CoRR, abs/1705.10694, 2017.

[17] V. Grancharov, D. Y. Zhao, J. Lindblom and W. B. Kleijn, "Low-Complexity, Nonintrusive Speech Quality Assessment," in *IEEE Transactions on Audio, Speech, and Language Processing*, vol. 14, no. 6, pp. 1948-1956, Nov. 2006. doi: 10.1109/TASL.2006.883250